\documentstyle[a4]{article}

\newcommand{\oo}{\"{o}}

\renewcommand{\d}{\mbox{\rm d}}
\def\lsim{\mathrel{\rlap{\lower4pt\hbox{\hskip1pt$\sim$}}
    \raise1pt\hbox{$<$}}}         
\def\gsim{\mathrel{\rlap{\lower4pt\hbox{\hskip1pt$\sim$}}
    \raise1pt\hbox{$>$}}}         

\def\title#1{
{\center\LARGE\bf{#1}\endcenter}}
\def\author#1{\vskip 1.5em{\center\large#1\endcenter}}
\def\affil#1{\vspace*{-1.0ex}{\topsep0pt\center{\topsep0pt\small\em #1}
\endcenter}}
\def\and{\vspace*{1.0ex}{\topsep0pt\center and\endcenter}}
\def\abstract{\vskip 1.5em\small{\center\large\bf Abstract\vspace{-.5em}
\vspace{0pt}\endcenter}\quotation}

\begin{document}


\newpage\null\thispagestyle{empty}
\rightline{PAR/LPTHE/95-12}
\rightline{UUITP-3/95}

\title{WIMP Mass Determination \\ with Neutrino Telescopes}

\author{Joakim~Edsj{\oo}\footnote{E-mail address: edsjo@teorfys.uu.se}}

\affil{\small \em Department of Theoretical Physics, Uppsala University,\\  Box
803, S-751 08 Uppsala, Sweden}

\and\author{Paolo~Gondolo\footnote{Postal address: LPTHE, Universit\'{e} de
Paris
VI \& VII, Tour 14--24,  5$^{\mbox{e}}$ \'etage, 2 place Jussieu, F-75251
Paris, France. E-mail address: gondolo@lpthe.jussieu.fr}}

\affil{\small \em Universit\'e Pierre \& Marie Curie, Paris VI\\ Universit\'e
Denis Diderot, Paris VII \\ Physique Th\'eorique et Hautes Energies\\ Unit\'e
associ\'ee au CNRS D 0280}

\begin{abstract}
\smallskip
Weakly-interacting massive particles (WIMPs) annihilating in the center of the
Sun or the Earth may give rise to energetic neutrinos which might be discovered
by astronomical neutrino detectors. The angular distribution of the
neutrino-induced muons is considered in detail via Monte Carlo simulations. It
is shown that large underground \v{C}erenkov neutrino telescopes might be able
to extract the WIMP mass from the width of the muon angular distribution.
\end{abstract}


\section{Introduction} \label{sec:intro}

Weakly-interacting massive particles (WIMPs) with masses in the GeV--TeV range
are among the leading non-baryonic candidates for the dark matter in our
galactic halo. One of the most promising methods for the discovery of WIMPs in
the halo is via observation of energetic neutrinos from annihilation of WIMPs
in the Sun \cite{neusun} and/or the Earth \cite{neuea}.  Through elastic
scattering with the atomic nuclei in the Sun or the Earth, a WIMP from the halo
can lose enough energy to remain gravitationally trapped \cite{trap}. Trapped
WIMPs sink to the core of the Sun or the Earth where they annihilate into
ordinary particles: leptons, quarks, gluons and -- depending on the masses --
Higgs and gauge bosons. Because of absorption in the solar or terrestrial
medium, only neutrinos are capable of escaping to the surface. Most WIMP
candidates -- among them the supersymmetric candidate, the neutralino -- do not
annihilate into neutrinos directly \cite{Goldberg}. Nevertheless energetic
neutrinos are eventually produced via hadronization and/or decay of the
annihilation products. These energetic neutrinos may be discovered by
astronomical neutrino detectors.

Trapped WIMPs are (to a good approximation) in thermal equilibrium with the
core of the Sun and/or the Earth. The radial extension of the WIMP annihilation
region is a function of the WIMP mass \cite{GrSe,Gould87}, heavier WIMPs lying
deeper in the core. This led Gould \cite{Gould} to suggest that the WIMP mass
might be inferred from the angular size of the annihilation region.

In this letter, we consider \v{C}erenkov neutrino telescopes. They consist of
large underground arrays of photo-multipliers to detect the \v{C}erenkov light
emitted by muons generated in charged-current interactions of neutrinos with
the medium surrounding the detector. Underground \v{C}erenkov detectors,
originally built to search for proton decay, have already started to explore
(and constrain) WIMP dark matter candidates \cite{WIMPbounds}.

Here we want to study if \v{C}erenkov neutrino telescopes currently
planned or under construction \cite{newCerenkov} might realistically expect to
be able to extract the WIMP mass from the muon angular distribution (once
measured). We include the uncertainties in the determination of the neutrino
direction due to the neutrino-muon scattering angle in charged-current
interactions, to multiple Coulomb scattering of the muon on its way to the
detector and to an intrinsic angular resolution in the determination of the
direction of the muon track.


\section{Annihilation channels and muon fluxes} \label{sec:achann}

WIMPs trapped in the core of the Sun and/or Earth can annihilate to a
fermion-antifermion pair, to gauge bosons, Higgs bosons and gluons
($\chi\chi \to \ell^+\ell^-$, $q\bar{q}$, $gg$, $q\bar{q}g$, $W^+W^-$,
$Z^0Z^0$, $Z^0H^0$, $W^{\pm}H^{\mp}$, $H^0H^0$). These annihilation products
will hadronize and/or decay, eventually producing high energy muon neutrinos.

Edsj\oo\ \cite{Edpre} reconsidered the whole chain of processes from the
annihilation products in the core of the Sun or the Earth to detectable muons
at the surface of the Earth.  He performed a full Monte Carlo simulation of the
hadronization and decay of the annihilation products using {\sc Jetset} 7.3
\cite{Jetset}, of the neutrino interactions on their way out of the Sun and of
the charged-current neutrino interactions near the detector using {\sc Pythia}
5.6 \cite{Jetset}, and finally of the multiple Coulomb scattering of the muon
on its way to the detector using distributions from Ref.~\cite{PDG}.

With respect to previous calculations \cite{RS,neuprod}, the Edsj\oo\
Monte Carlo treatment of the neutrino propagation through the Sun
bypasses simplifying assumptions previously made, namely neutral
currents are no more assumed to be much weaker than charged currents
and energy loss is no more considered continuous. In the new
treatment, the neutrino energy spectrum at the surface of the Sun is
obtained as follows. The thickness of the Sun and the neutrino mean
free path determine the probability of neutrino-nucleus interactions.
Each interaction is randomly chosen to be a charged-current
interaction, in which case the neutrino is considered absorbed, or a
neutral-current interaction, in which case the neutrino is degraded in
energy according to distributions in {\sc Pythia} 5.6. The procedure
is continued until the neutrino has reached the surface of the
Sun. The resulting neutrino spectrum differs significantly from
previous calculations only in the high energy tail.  But from this
high energy tail comes the most important contribution to the muon
flux in \v{C}erenkov neutrino detectors.\footnote{We remind that this
is so because the muon flux is the product of the neutrino flux by the
charged-current cross section and the muon range, and both of these
are proportional to the neutrino energy.} Hence at a WIMP mass of 1500
GeV (50 GeV) Edsj\oo\ finds a muon flux 20\% (5\%) higher than that
obtained by Ritz and Seckel \cite{RS}.

For more details on Edsj\oo\ results, we refer the reader to
Ref.~\cite{Edpre}.  In the following, we rely on his results obtained
by simulating $10^5$ WIMP annihilation events per annihilation channel and
WIMP mass.


\begin{figure}
  \vspace{7.0cm}
  \includegraphics{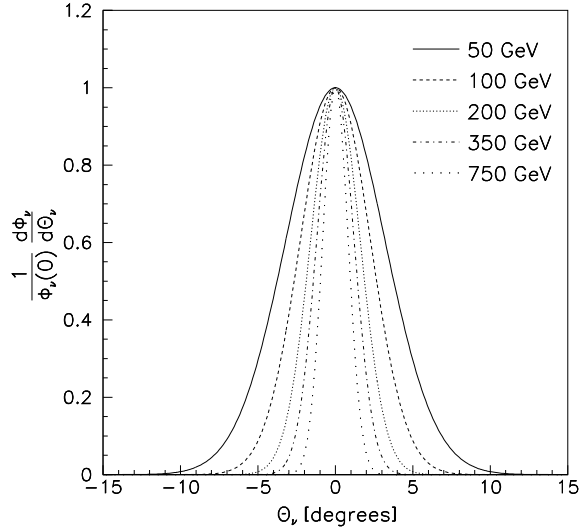}
  \caption{\em Projected angular distributions of WIMP-generated neutrinos
  from  the Earth for WIMP masses of (solid line) 50 GeV, (dashed line) 100
  GeV, (dotted line) 200 GeV, (dash-dotted line) 350 GeV and
  (wide dotted line)
  750 GeV. The analogous distributions from  the Sun are simply a
  narrow peak at $\theta_{\nu} = 0$.}
  \label{fig:aprof}
\end{figure}

\section{Annihilation profiles} \label{sec:aprof}

The annihilation region in the Sun can be regarded as point-like, its
angular size being $\lsim 0.005^{\circ}$ for the WIMP masses we are
interested in, $m \gsim 10$ GeV\@.  For the Earth, on the
contrary, the angular extension of the annihilation region is
non-negligible and decreases as one over the square root of the WIMP
mass \cite{GrSe}. In fact, the annihilation rate per unit volume at a
distance $r$ from the center of the Earth is proportional to the
square of the WIMP number density $n(r)$. The latter may be written as
\begin{equation} n(r) = n(0)
  \mbox{e} ^ { - r^2 / 2 r_{\chi}^2 } ,
\end{equation} with
\begin{equation}
  r_{\chi} = \left[ \frac{3 k T}{4 \pi G \rho m} \right]^{1/2}
  \simeq \frac{ 0.56 R_{\oplus} }{ \sqrt{ m/\mbox{GeV} } }.
\end{equation}
Here we have taken the radius of the Earth $R_{\oplus} \simeq $ 6400
km, the central Earth temperature $T \simeq 6000$ K and the central
Earth density $\rho \simeq 13$ g cm$^{-3}$\@. For the WIMP masses we
are interested in, it is a very good approximation to consider a
constant Earth density in the region where WIMPs are concentrated. For
an observer close to the surface of the Earth the angular distribution
of the neutrinos generated in WIMP annihilations results
\begin{equation} \label{eq:aprof}
  \frac{ \d \Phi_{\nu} } { \d \varphi_{\nu} \d \cos \theta_{\nu} }
  \propto \mbox{e}^{ - a^2 \sin^2 \theta_{\nu}}
  \mbox{erf}(a \cos \theta_{\nu})
\end{equation}
where
$\theta_{\nu} $ is the angle between the neutrino direction and  the center of
the Earth, $\varphi_{\nu}$ is the associated azimuthal angle,
$ a = {R_{\oplus}} / {r_{\chi}} $ and erf is the error function.

This expression simplifies in our case, $ m \gsim 10 $ GeV, for which
$\theta_{\nu}$ is typically smaller than $15^\circ$. In this case the
$\theta_{\nu}$-distribution is simply approximated as a gaussian in the
transverse plane, the plane orthogonal to the directions of either the Sun or
the Earth centers, \begin{eqnarray} \label{eq:aprofg}
\frac{\d\Phi_{\nu}}{\d\theta_x \d\theta_y} \propto  \mbox{e}^{- a^2 (
\theta_x^2 + \theta_y^2 ) }, \end{eqnarray} where $\theta_x$ and $\theta_y$
have obvious meaning.

To reduce the fluctuations due to limited statistics it is more  convenient to
consider the projected distribution in $ \theta_x $. From Eq.~(\ref{eq:aprofg})
we find that the projected distribution can be considered gaussian,
\begin{equation} \frac{\d\Phi_{\nu}}{\d\theta_x} \simeq \Phi_{\nu}(0)
\mbox{e}^{- a^2 \theta_x^2 }, \end{equation} with root mean square  value
\begin{equation} \label{eq:thrms} \theta_{\nu}^{rms} \simeq \frac{1}{\sqrt{2}}
\frac{r_{\chi}}{R_{\oplus}} \mbox{rad} \simeq   \frac{23^{\circ}}{\sqrt{
m/\mbox{GeV}}} \qquad ( m \gsim 10 \mbox{ GeV}) . \end{equation}

Projected angular distributions of the neutrino flux from the Earth
are shown in Fig.~\ref{fig:aprof} for WIMP masses between 50 and 750
GeV. The distribution width decreases for increasing WIMP masses. Note
that these distributions are independent of the neutrino energy
spectrum and of the specific annihilation channel. The analogous
distributions for the Sun are simply narrow peaks at $ \theta_\nu = 0
$.


\begin{figure}
  \vspace{7.0cm}
  \includegraphics{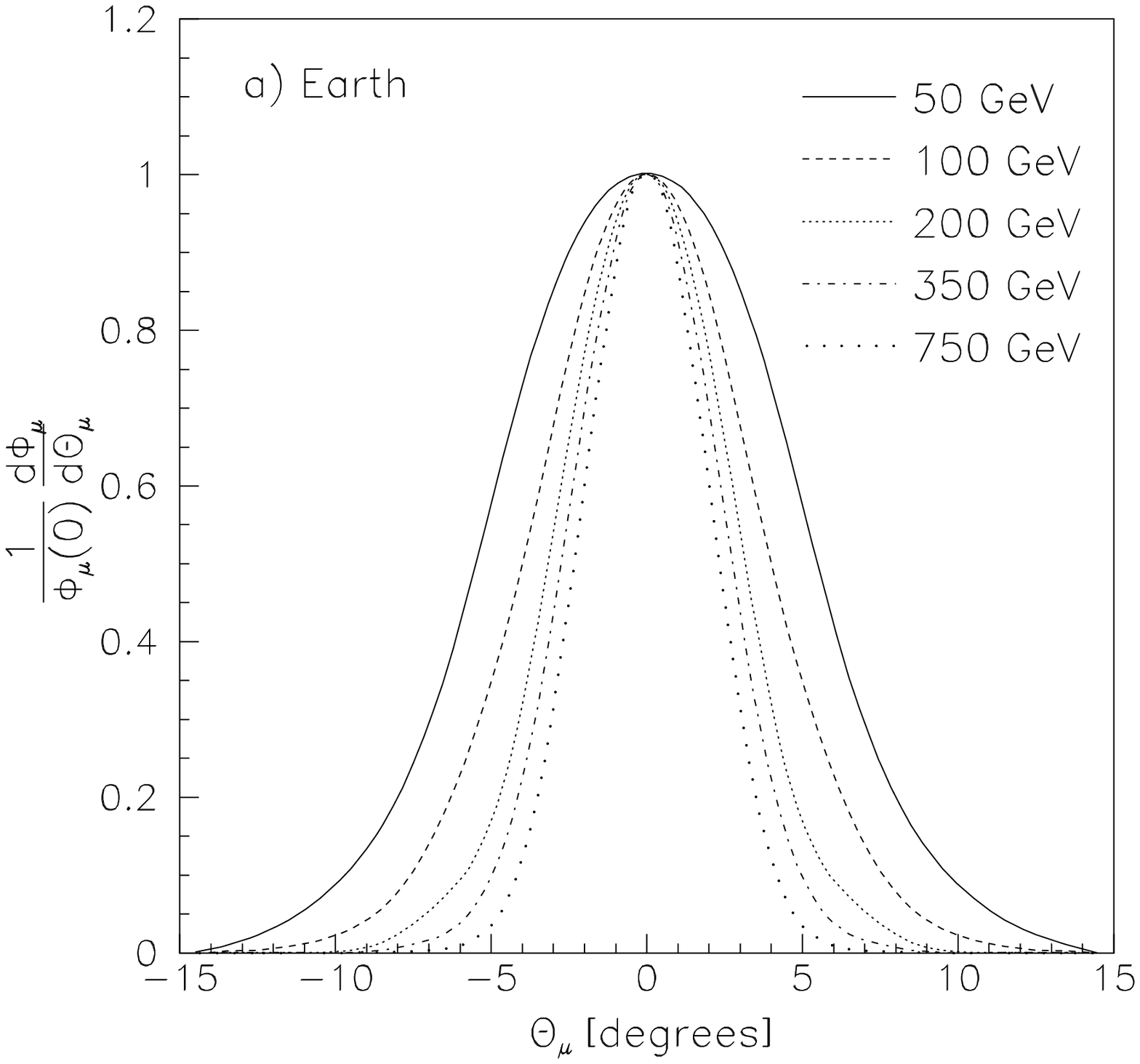}
  \includegraphics{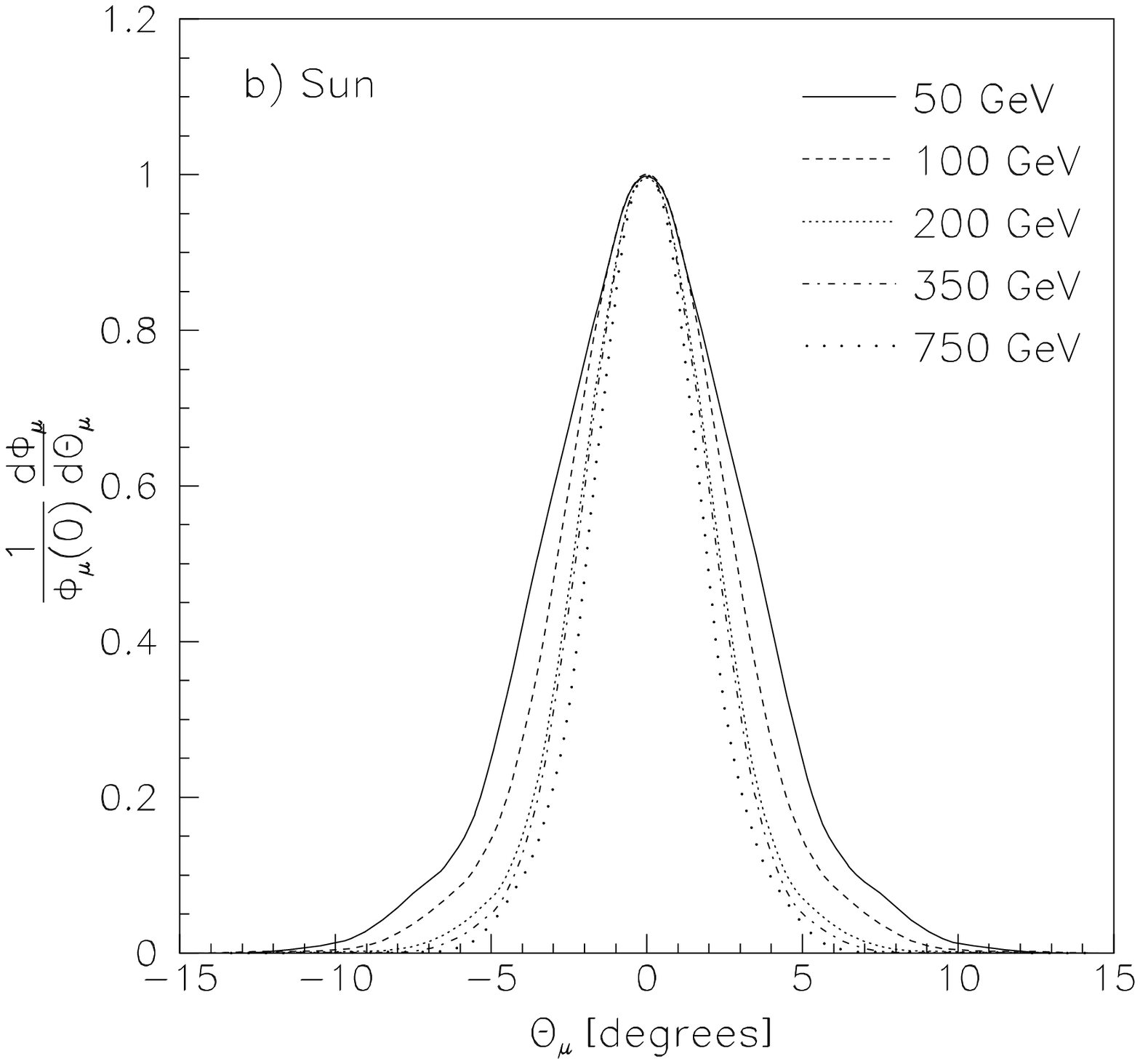}
  \caption{\em Projected angular distributions of neutrino-induced muons from
  WIMP annihilations in  (a) the Earth and (b) the Sun for WIMP masses of
  (solid line) 50 GeV, (dashed line) 100 GeV, (dotted line) 200 GeV,
  (dash-dotted line) 350 GeV and (wide dotted line) 750 GeV\@. The
  distributions are
  shown for hard channels ($W^+ W^-$ for 100--750  GeV and $\tau^+ \tau^-$ for
  50 GeV), with a detector muon threshold $E_{\mu}^{th} = 10 $ GeV and a
  detector angular resolution $\theta_{det}=1.4^\circ$.}
  \label{fig:wwtheta}
\end{figure}

\section{Muon angular distributions} \label{sec:adist}

In \v{C}erenkov neutrino telescopes it is not possible to measure the
angular distribution of the neutrinos directly since it is the muon
produced in charged-current interactions that can be detected.  The
direction of the neutrino is somewhat lost because of two effects: (1)
the muon produced in a charged-current interaction exits at an angle
$\theta_{CC}$ with respect to the incoming neutrino and (2) the same
muon undergoes multiple Coulomb scattering on its way to the detector,
changing direction by an angle $\theta_{Coul}$.  Both angles are
approximately gaussian in the transverse plane (at least in the
central region, for $ \theta_{CC} $ has non-gaussian tails) with root
mean square values
\begin{equation} \label{CCrms}
  \theta_{CC}^{rms}  \simeq  \frac{ 19^{\circ} } { \sqrt{E_{\nu}/\mbox{GeV}} }
  \hspace{1.5cm} \mbox{($E_{\mu}>10$ GeV)}
\end{equation}
and
\begin{equation} \label{Coulrms}
  \theta_{Coul}^{rms}  \simeq  \frac{
  3.1^{\circ} }{ \sqrt{E_{\mu}/\mbox{GeV}}},
\end{equation}
where the first
relation is obtained by simulations with {\sc Pythia 5.6} and the second
relation is from Ref.~\cite{PDG}\@. Notice that both
angles get smaller with increasing neutrino (and muon) energy.

There is an additional uncertainty coming from the reconstruction of the muon
track. Each neutrino telescope has an intrinsic angular resolution in
determining the direction of the muon. We assume that the error in its
determination is normally distributed with root mean square value
$\theta_{det}$, typically of the order of $1^\circ$.

In Fig.~\ref{fig:wwtheta} we plot the muon angular distributions for
hard channels in the Earth and in the Sun, obtained from the full
Edsj\"o simulations in \cite{Edpre}. These distributions are
representative of any hard neutrino spectrum. For a neutrino spectrum
to be hard it is not necessary that it is dominated by a hard channel,
like $W^+W^-$, $Z^0Z^0$ and $\tau^+ \tau^-$. Because of the
previously-mentioned importance of the high energy tails, it suffices
that the branching ratio into hard channels is greater than $\sim
10$\%. Softer neutrino spectra, {\it e.g.\/} those from the $b\bar{b}$
and $H^0H^0$ channels, give rise to wider angular distributions. This
is due to the energy dependence of $\theta_{CC}^{rms}$ and
$\theta_{Coul}^{rms}$.  Note the difference between the neutrino
(Fig.~\ref{fig:aprof}) and muon (Fig.~\ref{fig:wwtheta}) angular
distributions: charged-current interactions and multiple Coulomb
scattering make the width dependence on WIMP mass stronger.


\section{WIMP mass determination}

Information on the WIMP mass might be extracted from the width of the
distribution in the transverse plane. With limited statistics it might be
convenient to project this two-dimensional distribution onto the $\theta_x$
axis. As a measure of the width we choose the full width half maximum $
\theta_{FWHM} $ of the projected distribution. The FWHM is not sensitive to the
non-gaussian tails of the distribution, which reflect the non-gaussian tails in
$ \theta_{CC} $. Moreover, as long as the detector resolution $
\theta_{det} $ is small with respect to $ \theta_{FWHM} $, it should be
relatively easy to extract the FWHM even in the presence of a background (muons
and neutrino-induced muons from cosmic ray interactions in the atmosphere).

\begin{figure}
\begin{picture}(16.01,7.5)(0,0)
  \put(1.9,6.7){\large a)}
  \put(10.15,6.7){\large b)}
  \includegraphics{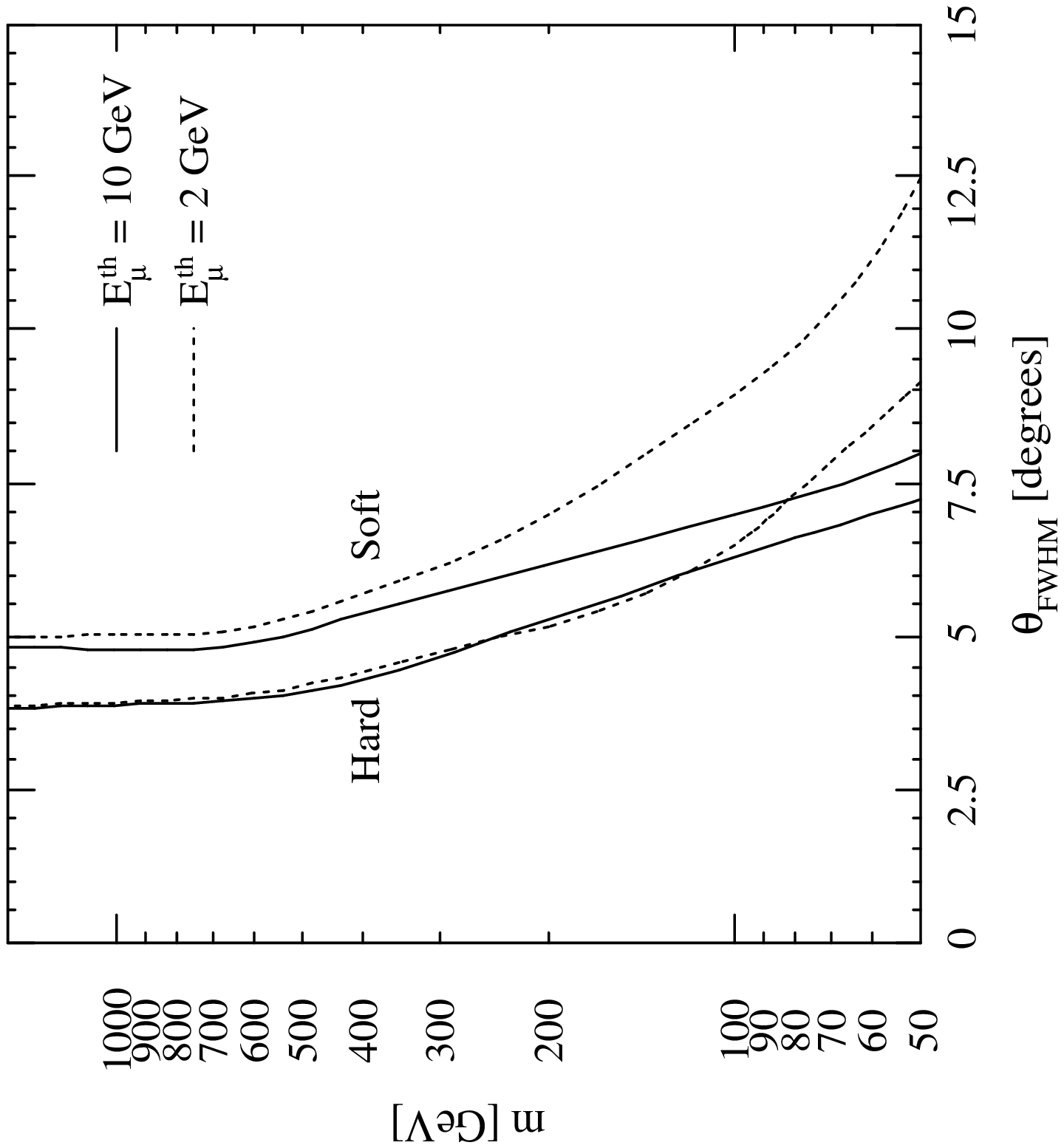}
  \includegraphics{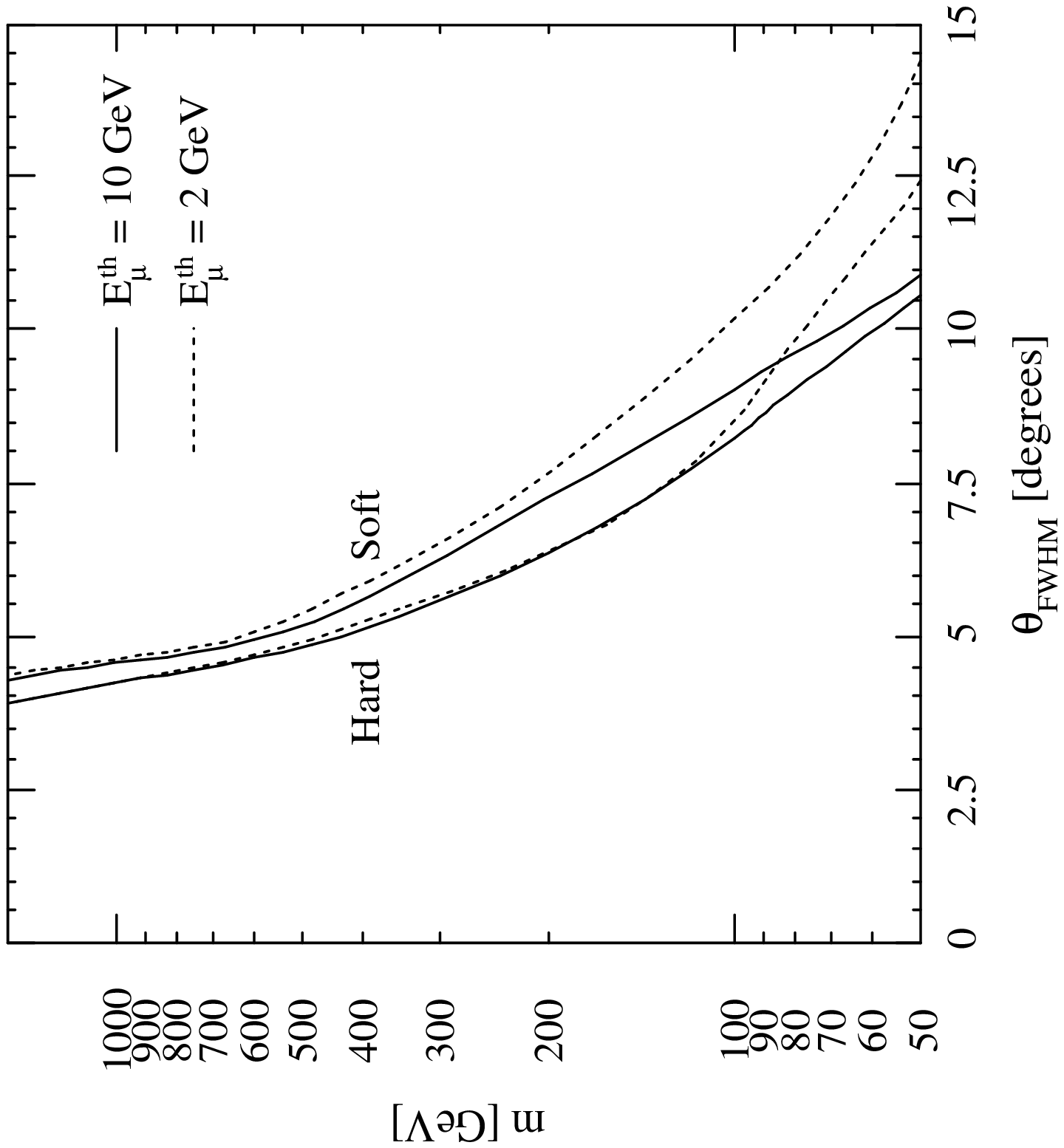}
\end{picture}
\begin{picture}(16.01,7.5)(0,0)
  \put(1.9,6.7){\large c)}
  \includegraphics{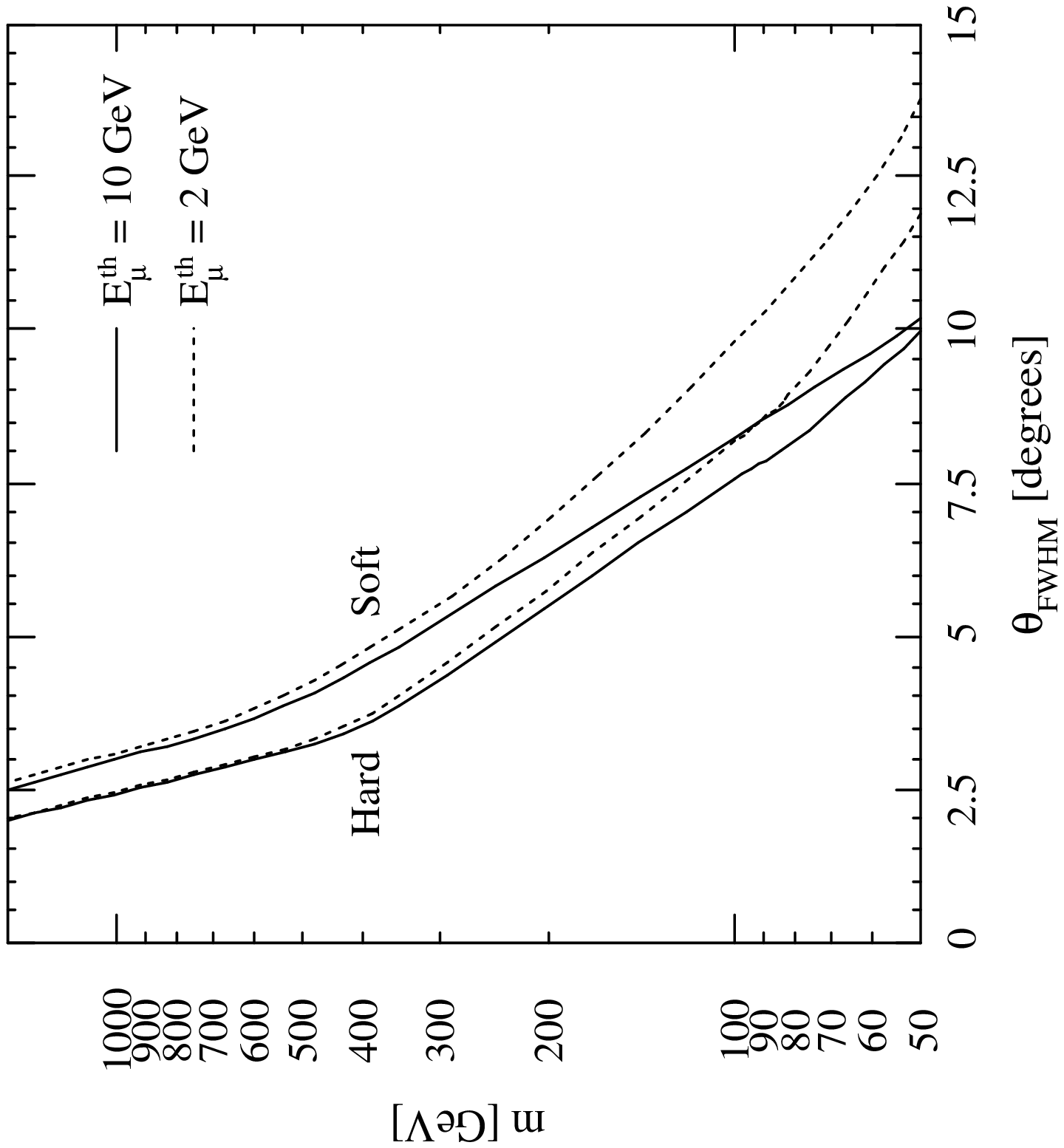}
\end{picture}
  \caption{\em WIMP mass $m$ versus full width half maximum $\theta_{FWHM}$ of
  the neutrino-induced muon distribution for soft ($b
  \bar{b}$)  and hard ($W^+ W^-$ and $\tau^+ \tau^-$) channels. The solid lines
  correspond to  $E_{\mu}^{th} = 10 \mbox{ GeV}$ and the dashed lines
  to  $E_{\mu}^{th} = 2 \mbox{ GeV}$. a) For Sun-bound WIMPs with a
  detector angular resolution of  $\theta_{det} = 1.4^\circ$. b) For
  Earth-bound WIMPs with a detector angular resolution of $\theta_{det}
  = 1.4^\circ$. c) For Earth-bound WIMPs with a perfect angular resolution. }
  \label{fig:thmfwhm}
\end{figure}

In Figs.~\ref{fig:thmfwhm}a-c we show the
dependence of the WIMP mass on the full width half maximum for some
representative cases. We present the soft $b\bar{b}$-channel and the  hard
$W^+W^-$- and $\tau^+ \tau^-$-channels for the Earth and the Sun for two
different muon energy thresholds, $ E_{\mu}^{th} = $ 2~GeV and $ E_{\mu}^{th}
= $ 10~GeV\@.
Figs.~\ref{fig:thmfwhm}a and \ref{fig:thmfwhm}b include a detector
angular resolution $\theta_{det}=1.4^\circ$\@. For the sake of comparison,
Fig.~\ref{fig:thmfwhm}c shows the  ideal case of a perfect angular resolution.
We have checked that all distributions are indeed well approximated  by
gaussians in the central regions.

We see that the detector angular resolution is the limiting factor for
the mass determination of heavy WIMPs ($ m \gsim 400 $ GeV)\@. For
lighter WIMPs, it seems promising to infer their mass from the Earth
muon distributions. We remind that this is also the mass range in
which the signal from the Earth is expected to be significant
\cite{Gould87}\@. The WIMP mass could also be extracted from the Sun
muon distributions provided the detector muon energy threshold is
low. In fact, the width of the angular distribution for the Sun is
dominated by charged-current scattering, which acts as a mass
spectrometer in diffusing neutrinos according to their energies and so
according to the WIMP mass.

{} From Figs.~\ref{fig:thmfwhm}a, \ref{fig:thmfwhm}b and analogous ones,
approximate relations between the error on the WIMP mass and the error
in $\theta_{FWHM}$ may be obtained. For example, for $ E_{\mu}^{th} =
$ 10 GeV, $ \theta_{det} = 1.4^\circ$ and $m \lsim 300$ GeV, we read
from Fig.~\ref{fig:thmfwhm}b
\begin{equation}
\Delta \log m \simeq 0.15 + 0.15 \Delta \theta_{FWHM}\label{eq:dm}
\end{equation}
with $ \Delta \theta_{FWHM} $ in degrees.  The first term represents
our ignorance on the actual annihilation channel and the second term
comes from the slope of the $m$ -- $\theta_{FWHM}$ curve. For $\Delta
\theta_{FWHM} \sim 1^\circ$, relation~(\ref{eq:dm}) gives an
uncertainty on the WIMP mass of roughly a factor of 2.0.

\begin{figure}
  \vspace{7.0cm}
  \includegraphics{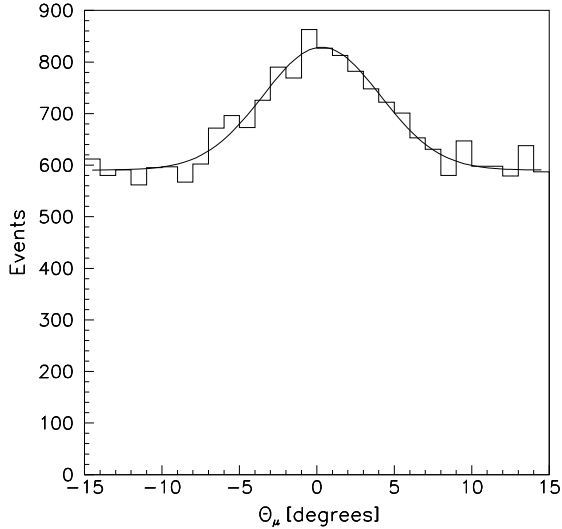}
  \caption{\em An example of mass determination from a projected muon angular
  distribution. The simulated histogram includes the expected background
  in one year of exposure with a 1 km$^2$ detector and a typical signal
  from annihilation of 100 GeV WIMPs in the Earth.
  The solid line is a fit of a gaussian plus a constant. }
  \label{fig:ex}
\end{figure}

We present now an example of mass determination for Earth-bound
WIMPs. We consider one year of exposure of a 1 km$^2$ detector with
a muon energy threshold of 10 GeV\@. The atmospheric background in the
direction of the center of the Earth is expected to be 20 muons per
square degree \cite{Gai84}. We choose a WIMP mass of 100 GeV and
generate a signal of 2000 muons, a reasonable number for
supersymmetric models with neutralinos of this mass. In this way we
obtain a simulated muon angular distribution in a $15^\circ \times
15^\circ$ region centered towards the center of the Earth. We then
analyze these simulated data.  We project the muon distribution onto
the $\theta_x$ axis and obtain the histogram shown in
Fig.~\ref{fig:ex}. By fitting a gaussian plus a constant to this
histogram we obtain a full width half maximum $\theta_{FWHM} =
8.9^\circ \pm 0.7^\circ$. From Fig.~\ref{fig:thmfwhm}b we then read the
WIMP mass range corresponding to the fitted FWHM range,
$m=90^{+50}_{-25}$ GeV\@.  We are satisfied that the mass range
obtained contains the original WIMP mass.  The uncertainty is
approximately a factor of 1.5, in agreement with the approximate
relation Eq.~(\ref{eq:dm}). One might worry about the size of the
uncertainty, but we believe that even a rough determination of the
WIMP mass would be of enormous importance. We are exploring ways to
reduce this uncertainty.

In favorable cases, many muon neutrinos might be detected from both
the Earth and the Sun. Having both angular distributions, the angular
smearing due to charged-current interactions and multiple Coulomb
scattering would be directly represented by the Sun muon distribution
(recall that the annihilation region in the Sun can be considered
pointlike). This distribution might then be subtracted from the Earth
muon distribution, leaving the neutrino angular distribution from the
Earth. From it the WIMP mass could be obtained in a direct way via
Eq.~(\ref{eq:aprof}).  Notice that for heavy WIMPs ($m \gsim 100$ GeV)
one should correct for absorption of neutrinos on their way out of the
Sun, {\it e.g.\/} by specifying the shape of the neutrino energy
spectrum.


\section{Conclusions} \label{sec:concl}

WIMPs annihilating in the center of the Sun or the Earth may give rise
to a neutrino-induced muon flux in astronomical neutrino detectors.
The width of the muon angular distribution carries information on the
WIMP mass, because the size of the annihilation region, the
charged-current neutrino-nucleon scattering and the multiple Coulomb
scattering of the muons all depend on the WIMP mass.  Detailed Monte
Carlo simulations have been used to obtain the muon angular
distribution for WIMP annihilations in the Earth and in the Sun. It
has been shown that the WIMP mass can be inferred, for WIMPs lighter
than $\sim 400$ GeV, from the Earth distribution and, provided the
muon energy threshold is low ($\lsim 5$ GeV), also from the Sun
distribution.  This seems therefore a promising method of determining
the WIMP mass and we look forward for the detection of a WIMP signal
in neutrino telescopes.


\section*{Acknowledgments} \label{sec:ack}

We would like to thank L.~Bergstr\"{o}m for interesting comments.
J.~Edsj{\"o} is also grateful to L.~Bergstr{\"o}m and G.~Ingelman for
valuable discussions on the physics behind the simulations.
This work has been partially supported by the EC
Theoretical Astroparticle Network under contract No.~CHRX-CT93-0120
(Direction G\'en\'erale 12 COMA).


\end{document}